\documentclass[twocolumn,showpacs,amsmath,amssymb,amsfonts,floatfix,prl,aps]{revtex4}
\usepackage{graphicx}
\usepackage{color}

\begin{document}
\title{Enhanced excitonic effects in the energy loss spectra of LiF and Ar at large momentum transfer}
\author{S. Sharma$^1$}
\email{sharma@mpi-halle.mpg.de}
\author{J. K. Dewhurst$^1$}
\author{A. Sanna$^1$}
\author{A. Rubio$^2$}
\author{E. K. U. Gross$^1$}
\affiliation{1 Max-Planck-Institut f\"ur Mikrostrukturphysik, Weinberg 2, 
D-06120 Halle, Germany.}
\affiliation{2 Nano-Bio Spectroscopy group and ETSF Scientific Development Centre, Centro de F\'isica de 
Materiales CSIC-UPV/EHU-MPC and DIPC, Universidad del Pa\'is Vasco UPV/EHU, Avenida de Tolosa 72, E-20018 San Sebastian, Spain}
\date{\today}

\begin{abstract}
It is demonstrated that the bootstrap kernel [\onlinecite{sharma11}] for finite values of ${\bf q}$ crucially depends upon 
the matrix character of the kernel and gives results  of the same good quality as in the  ${\bf q} \rightarrow 0$ limit.
The bootstrap kernel is further used to study the electron loss as well as absorption spectra for Si, LiF and Ar for 
various values of ${\bf q}$. The results show that the excitonic effects in LiF and Ar are enhanced for values of ${\bf q}$ 
away from the $\Gamma$-point. 
The reason for this enhancement is the interaction between the exciton and high energy inter-band electron-hole transitions. 
This fact is validated by calculating the absorption spectra under the influence of an external electric field. The 
electron energy loss spectra is shown to change dramatically as a function of ${\bf q}$.
\end{abstract}

\maketitle

One of the promising routes for tailoring electro-magnetic interactions in systems is to make use
of excitons-- excitonic condensation\cite{keldysh,eisen}, exciton-plasmon interaction\cite{bondarev,bellessa} and strong coupling of excitons with other inter-band optical transitions.  
A lot of applications of excitonic manipulation have found their way into surface physics, nano-structures and nanotubes. Extended
solids, however, have not been treated as serious candidates for such effects. This is mainly due to the fact that excitonic effects are in
general (but not exclusively) stronger in lower dimensional systems\cite{dim}. 

As for solids, much of the attention of excitonic physics has been focused on the determination of optical absorption spectra (specially in the long 
wavelength ($q \rightarrow 0$) limit). Determination of optical absorption spectra in this limit requires an accurate description
of electron-hole effects, which in turn requires  a computationally expensive many-body treatment at the level of the Bethe-Salpeter equation 
(BSE)\cite{hanke78,onida95,onida02,rohlfing98,lask07,hahn05,bechstedt05,mula05}. 
Alternatively, this electron-hole physics can also be effectively treated using time-dependent density functional theory (TDDFT)\cite{tddft}. However,
there exist only a few  exchange-correlation (xc) kernels within TDDFT which are capable of accurately describing the excitonic effects 
arising from strong electron-hole interactions;
the computationally demanding but accurate nano quanta kernel\cite{reining02,marini03} which is derived from the BSE, 
the long-range corrected (LRC) kernel\cite{botti04-1,botti05,reining02} which has the form $\alpha/|q|^2$ (with $\alpha$ being a system dependent external parameter)
and the recently proposed bootstrap kernel\cite{sharma11}. The main feature which makes these kernels capable of accurately treating the 
absorption spectra (in $q \rightarrow 0$ limit) is their $1/|q|^2$ dependence\cite{ghosez97}, a feature necessary to capture the 
electron-hole physics.

In contrast to this the optical absorption spectra and the electron energy loss spectra (EELS) at finite 
values of ${\bf q}$ (away from the $\Gamma$-point), are known to be accurately treated\cite{botti07,hansi} by the
adiabatic local density approximation (ALDA)\cite{alda}, which does not have the $1/|q|^2$ dependence. 
This then raises an interesting question about the validity of the kernels which accurately treat the ${\bf q} \rightarrow 0$, 
for finite values of ${\bf q}$.

In the present work, the EELS and absorption spectra for Si, diamond, LiF and Ar
are calculated, using the bootstrap kernel\cite{sharma11}
implemented within the ELK code\cite{elk}. This choice is motivated by the fact that the bootstrap kernel is computationally not demanding and 
it does not depend upon any  system-dependent parameters. For prototypical materials with bound exlectron-hole pair, LiF and Ar, there exists 
a strong coupluing  between exciton with high energy inter-band transitions, which leads to strong excitonic effects away from the $\Gamma$-point. 
This coupling provides a useful handle to manipulate the excitonic physics, since 
inter-band transitions can easily be modified by changing external parameters like pressure and/or electric field.  

\begin{figure}[ht]
\begin{tabular}{c}
\includegraphics[width=\columnwidth,clip]{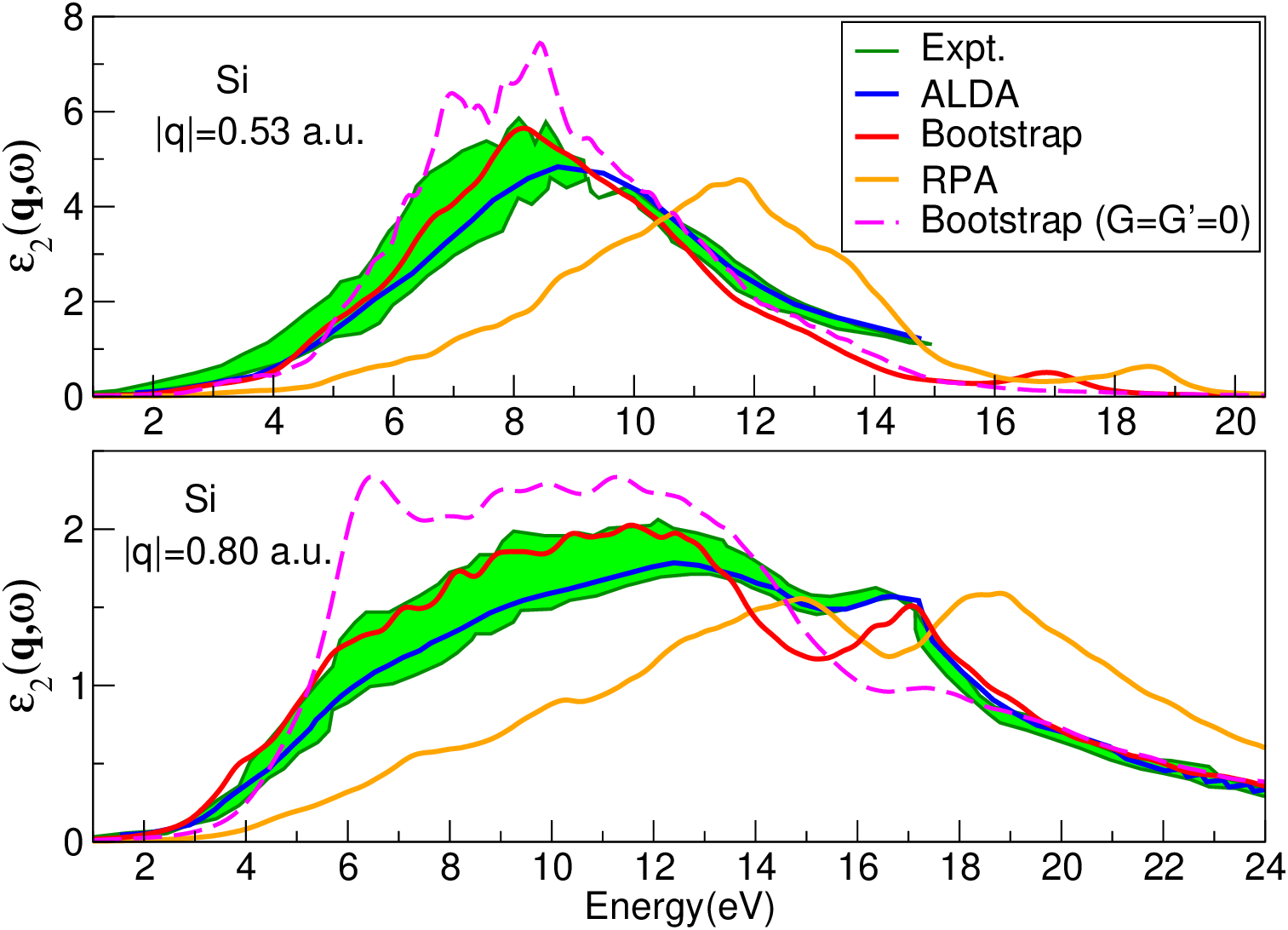} \\
\includegraphics[width=\columnwidth,clip]{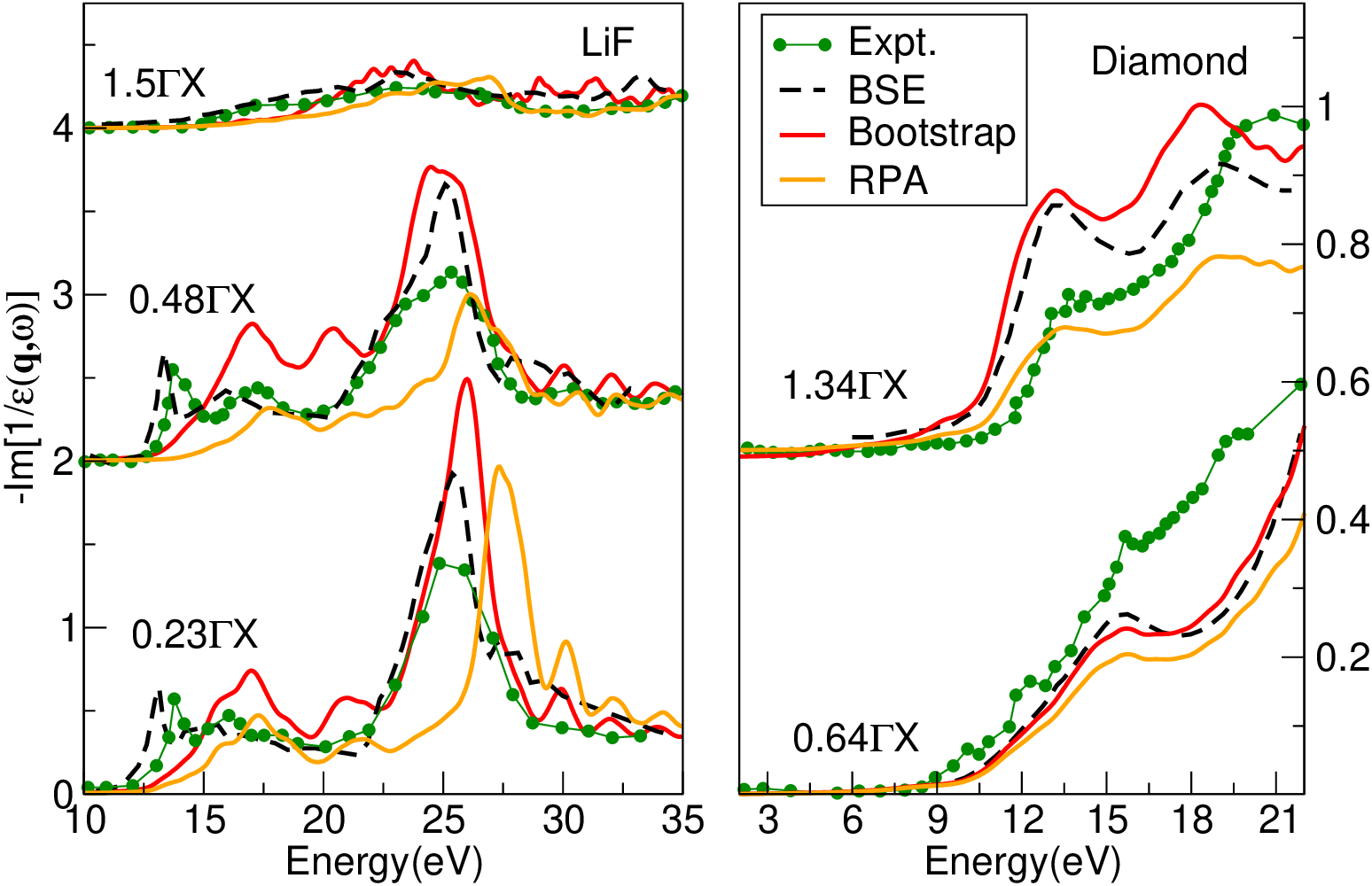}
\end{tabular}
\caption{Upper two panels-- imaginary part of the dielectric tensor as a function of energy in eV for Si. Results obtained using the bootstrap 
kernel, RPA and ALDA are compared with experimental data from Ref. \onlinecite{hansi}. The top panel contains results for $|{\bf q}|=0.53 a.u.$ and the
middle panel for $|{\bf q}|=0.80 a.u.$ parallel to the [111] direction. The bootstrap results are obtained in two different ways; (red line)
the kernel is a full matrix in the reciprocal space with $|{\bf G}|_{\rm max}=4 a.u.$ and (pink dotted line) only the head of
the kernel is used i.e. $|{\bf G}|_{\rm max}=0 a.u.$
The bottom panel contains results for the EELS, for different values of ${\bf q}$ (indicated in the figure) as a function of 
energy in eV for LiF and diamond. 
The results obtained using the bootstrap kernel are shown with full line, the experimental data (from Ref. \onlinecite{expt-lif}) with dots
and the BSE results (also taken from Ref. \onlinecite{expt-lif}) with dashed line.
The results for different values of ${\bf q}$ are shifted vertically for clarity. 
\label{si-lif}}
\end{figure}

The bootstrap xc-kernel within TDDFT for calculating linear response reads:
\begin{align}
 f^{\rm boot}_{\rm xc}({\bf q},\omega)=
 -\frac{\varepsilon^{-1}({\bf q},\omega=0)v({\bf q})}
 {\varepsilon_0^{00}({\bf q},\omega=0)-1}=
\frac{\varepsilon^{-1}({\bf q},\omega=0)}{\chi_0^{00}({\bf q},\omega=0)}
\end{align}
where, $v$ is the bare Coulomb potential, $\chi_0$ is the response function of the non-interacting Kohn-Sham system
and $\varepsilon_0({\bf q},\omega) \equiv 1-v({\bf q})\chi_0({\bf q},\omega)$
is the dielectric function in the random phase approximation (RPA). All these quantities are matrices in the basis of reciprocal 
lattice vectors ${\bf G}$.
With a view towards studying  the excitonic properties of solids for any value of ${\bf q}$, we first 
show that the bootstrap kernel for finite values of ${\bf q}$ leads to accurate results for the absorption spectra as well as
the EELS. This is demonstrated in Fig. \ref{si-lif} which compiles results for the dielectric response
and EELS of a small bandgap insulator, Si, a medium badgap insulator, diamond, and a large bandgap insulator, LiF. 
The choice of these materials is motivated by the abundance of available experimental data
and because they have been used as prototypical test cases for the influence of many-body corrections.

The upper two panels of Fig. \ref{si-lif} are presented the results for $\varepsilon({\bf q},\omega)$ of 
Si.
The results obtained using the bootstrap kernel are compared to the results obtained using the ALDA and to the experimental data. 
It is clear that as far as the absorption spectrum is concerned the bootstrap kernel (treated as a full matrix in reciprocal space) 
gives results in good agreement with experiments. 
Interestingly, for small frequencies the results obtained using the bootstrap kernel give an upper bound to the experimental data, while the 
results obtained using the ALDA give a lower bound. 
The RPA results (also shown in Fig. \ref{si-lif}), as expected, 
totally miss the excitonic physics, which in this case shows up as the shifting of the spectral weight to lower frequencies.  

In the ${\bf q} \rightarrow 0$ limit the ${\bf G}={\bf G'}=0$ component of f$_{xc}$ is the most 
important one and hence the bootstrap procedure can be thought of as a self-consistent method for obtaining the system dependent parameter 
$\alpha$ of the LRC. While, for finite values of ${\bf q}$ the matrix character of f$_{xc}$ is crucial and the bootstrap
kernel is significantly different from the LRC kernel. The importance of including higher ${\bf G}$ vectors in f$_{xc}$  
is demonstrated in Fig. \ref{si-lif}; results for Si show that the dielectric function obtained using only the 
${\bf G}={\bf G'}=0$ component of f$_{xc}$ are much higher in magnitude and in relatively poor agreement with the experimental data.

EELS corresponds to the negative of the  imaginary part of $\varepsilon_{00}^{-1}({\bf q},\omega)$, where 00 stands for the ${\bf G}={\bf G'}=0$ 
component of the dielectric tensor. The lower panel of Fig. \ref{si-lif} contains results for the EELS of LiF and diamond. 
Three different values of ${\bf q}$ in the $\Gamma-X$ direction are presented for LiF.  
Within the first BZ the experimental data and BSE results show three main peaks which are well reproduced by the bootstrap kernel. On going from 
0.23 to 0.48 $\Gamma X$ the plasmonic peak at 25 eV gets smaller in magnitude, a feature which is again well captured by the bootstrap kernel. 
Experiments, BSE and bootstrap results show that outside the first BZ (for ${\bf q}=1.50 \Gamma X$) EELS is highly suppressed.
These results indicate that the bootstrap kernel captures the change in  ${\rm -Im}[\varepsilon^{-1}]$ as a function of ${\bf q}$
very well. We note, however, that the magnitude of the peaks is
slightly overestimated by the bootstrap kernel and, for small energies, the peaks are blue shifted by $\sim$1 eV compared to experiment.
This shifting of the excitonic peaks to 
higher frequencies and the overestimation of their magnitude was also a feature of the absorption spectra in the long wavelength limit. 

For diamond, the magnitude of the EELS obtained using the bootstrap kernel is overestimated compared to the experiment. A similar 
overestimation is also seen in the BSE results\cite{expt-lif}. In fact, we find that the results obtained using the
bootstrap kernel are in very good agreement with the BSE results. As in the case of Si, for LiF and diamond as well, the RPA 
results are shifted to higher frequencies and are missing excitonic physics.

It is apparent from the examples above that the bootstrap results for values of 
${\bf q}$ away from the $\Gamma$-point, have the same good agreement with experiments as those in the 
${\bf q} \rightarrow 0$ limit.
With this in hand we can now use the bootstrap kernel to study the excitonic effects in various directions in the BZ. 
%

\begin{figure}[ht]
\begin{tabular}{c}
\includegraphics[width=\columnwidth,clip]{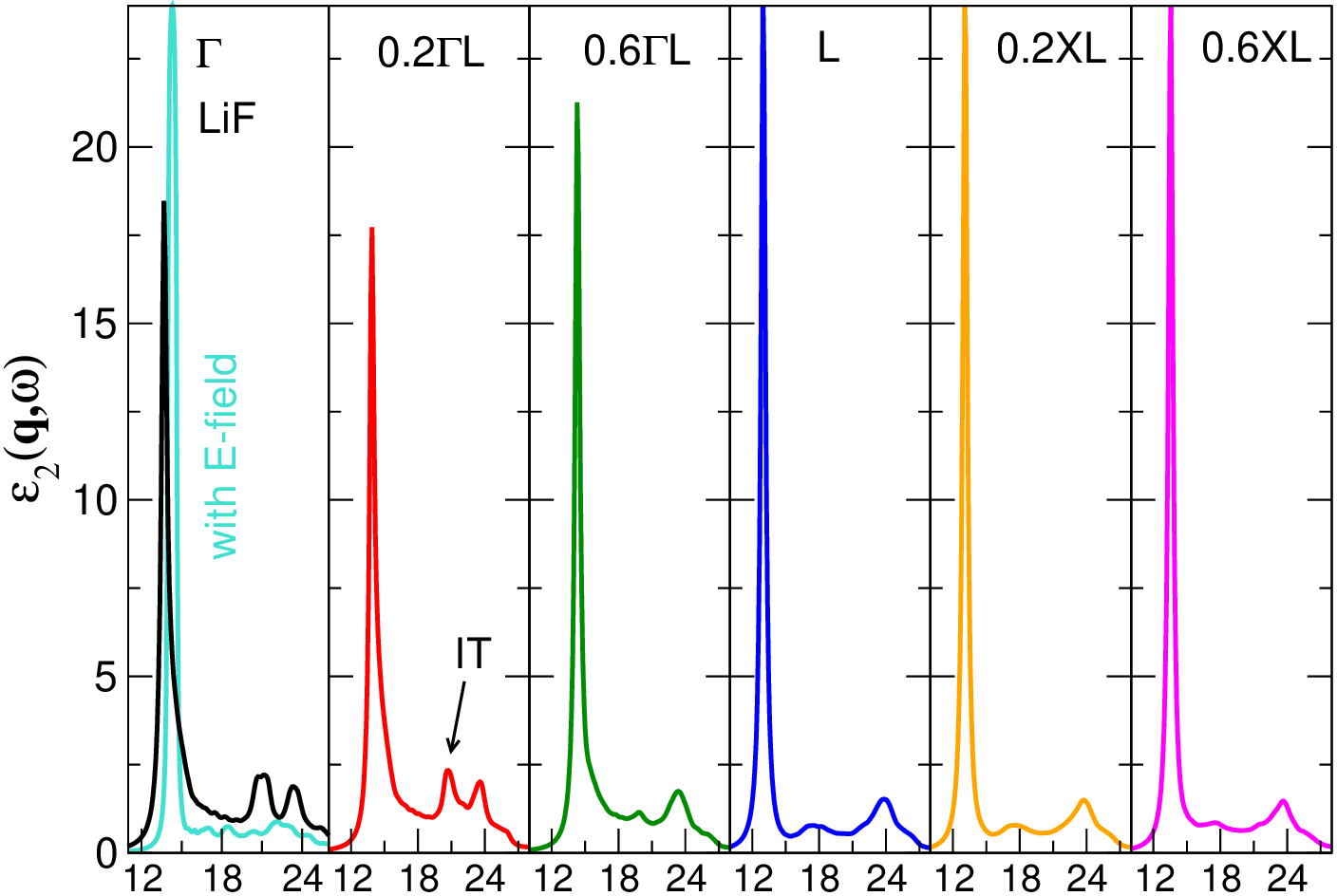} \\
\includegraphics[width=\columnwidth,clip]{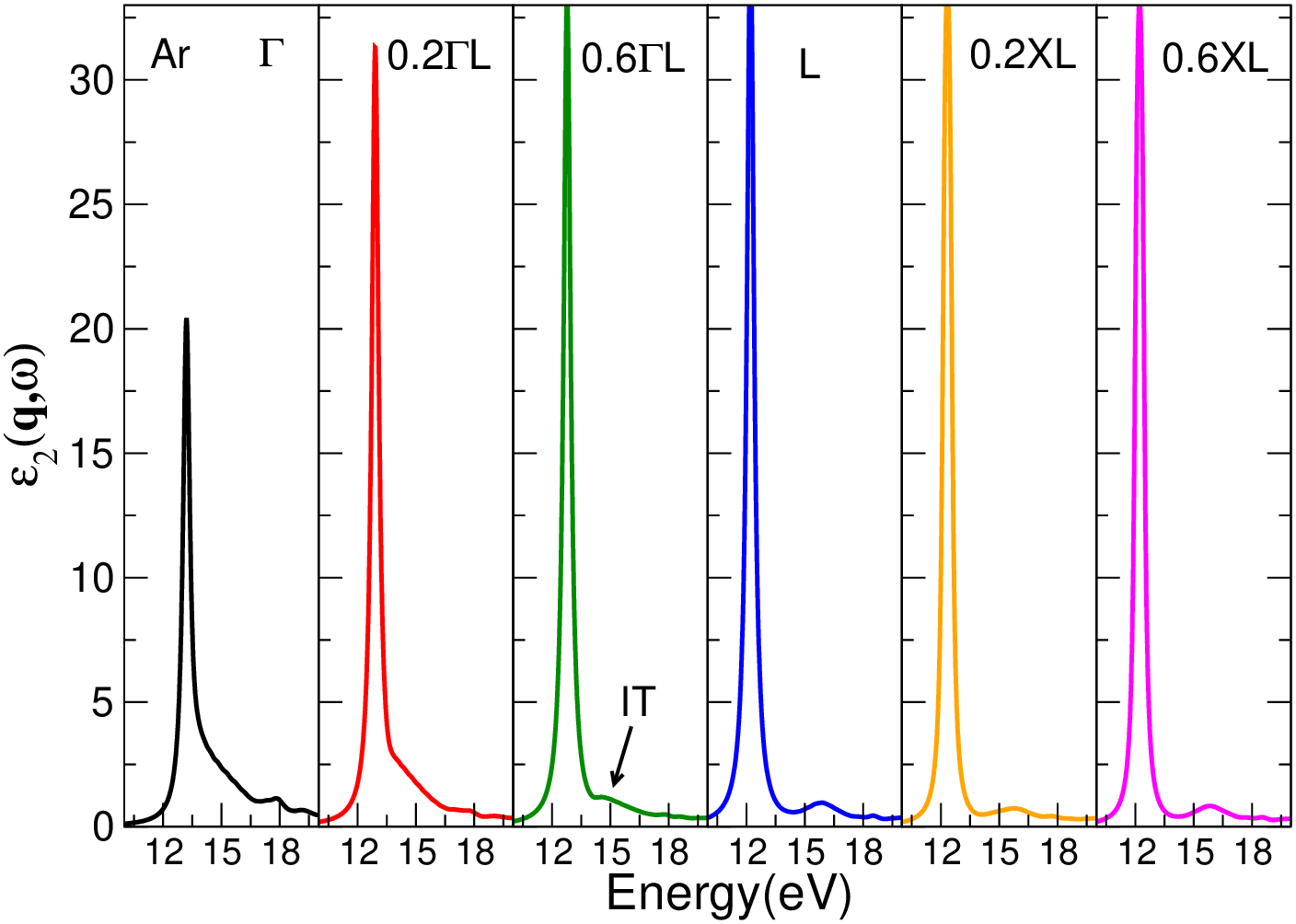}
\end{tabular}
\caption{Dielectric tensor obtained using the bootstrap kernel as a function of energy in eV. 
Upper panel contains results for LiF and the Lower panel for Ar in the $\Gamma-L-X$ direction. 
The upper-left most panel also contains results at the $\Gamma$-point for LiF exposed to an external electric field.  
\label{eps}}
\end{figure}

In  Fig. \ref{eps}  the results for $\varepsilon({\bf q},\omega)$ are shown for LiF and Ar for various values of ${\bf q}$. The upper panel 
contains results for LiF in the $\Gamma-L-X$ direction\cite{dirs}. It is immediately clear from this that the excitonic peak in 
LiF becomes stronger (spectral weight moves to lower frequency) away from the $\Gamma$-point; on going from $\Gamma$ to $L$, 
the excitonic peak ($\sim$13-14 eV) becomes larger in magnitude and at the same time the inter-band transition (IT) peak $\sim$20 eV 
diminishes and moves to lower frequencies (from 20 eV to 18 eV). In the $L-X$ direction the excitonic and IT peaks remain almost the same. 

Similar results are also seen for Ar; on going from $\Gamma$ to $L$, the spectral weight moves towards the excitonic 
peak ($\sim$12eV) which becomes stronger
and shifts to lower frequencies. This is accompanied by a steady diminishing of the IT peak around 18 eV.
Beyond $0.6 \Gamma L$, both the excitonic and the IT peaks are unchanged. 
The RPA for both LiF and Ar misses this excitonic physics for all values of ${\bf q}$. 

The above results point towards a strong coupling between the excitonic and the IT peaks; as the excitonic peak gains weight, the IT peak
diminishes. This observation could be used in the future to tune excitonic effects via manipulation of inter-band transitions. 
In order to validate this hypothesis we have performed calculations of the ${\bf q} \rightarrow 0$ absorption spectrum
for LiF in presence of an external electric field. This field is an artifice used to lift the degeneracy and split the bands.
This split in the bands causes the IT peak at 20 eV to diminish and move to lower 
energies. The excitonic peak gains weight (spectral weight moves to lower energy) and the results for $\varepsilon$ in 
the ${\bf q} \rightarrow 0$ limit closely resemble those of the $L$-point (see top-first panel of Fig. \ref{eps}). 
Such a tuning of the excitons has been performed before: a coupling of excitons and surface plasmons was used to 
enhance excitonic effects in low dimensional systems\cite{bellessa,bondarev,atta11}.

\begin{figure}[ht]
\begin{tabular}{c}
\includegraphics[width=\columnwidth,clip]{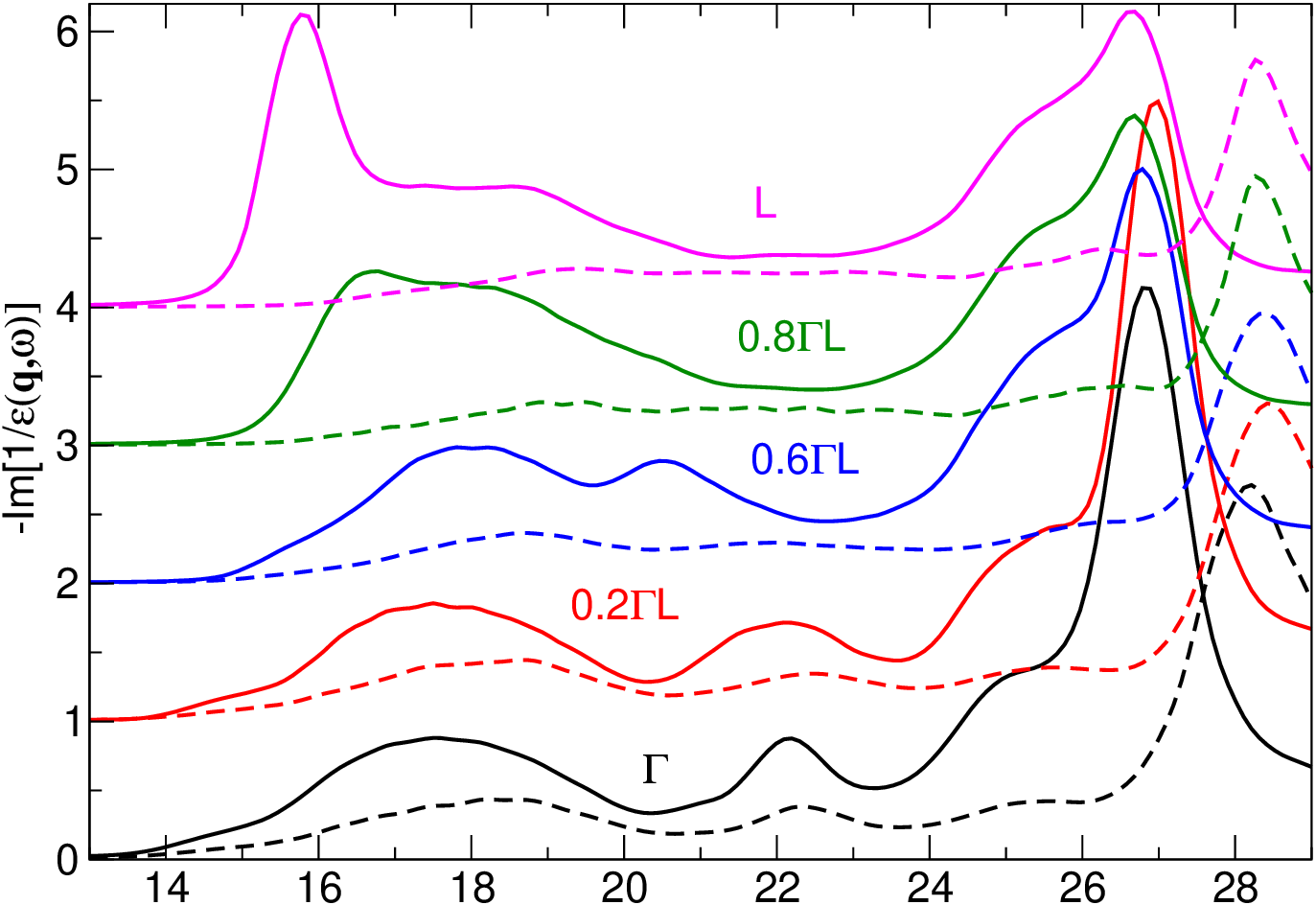} \\
\includegraphics[width=\columnwidth,clip]{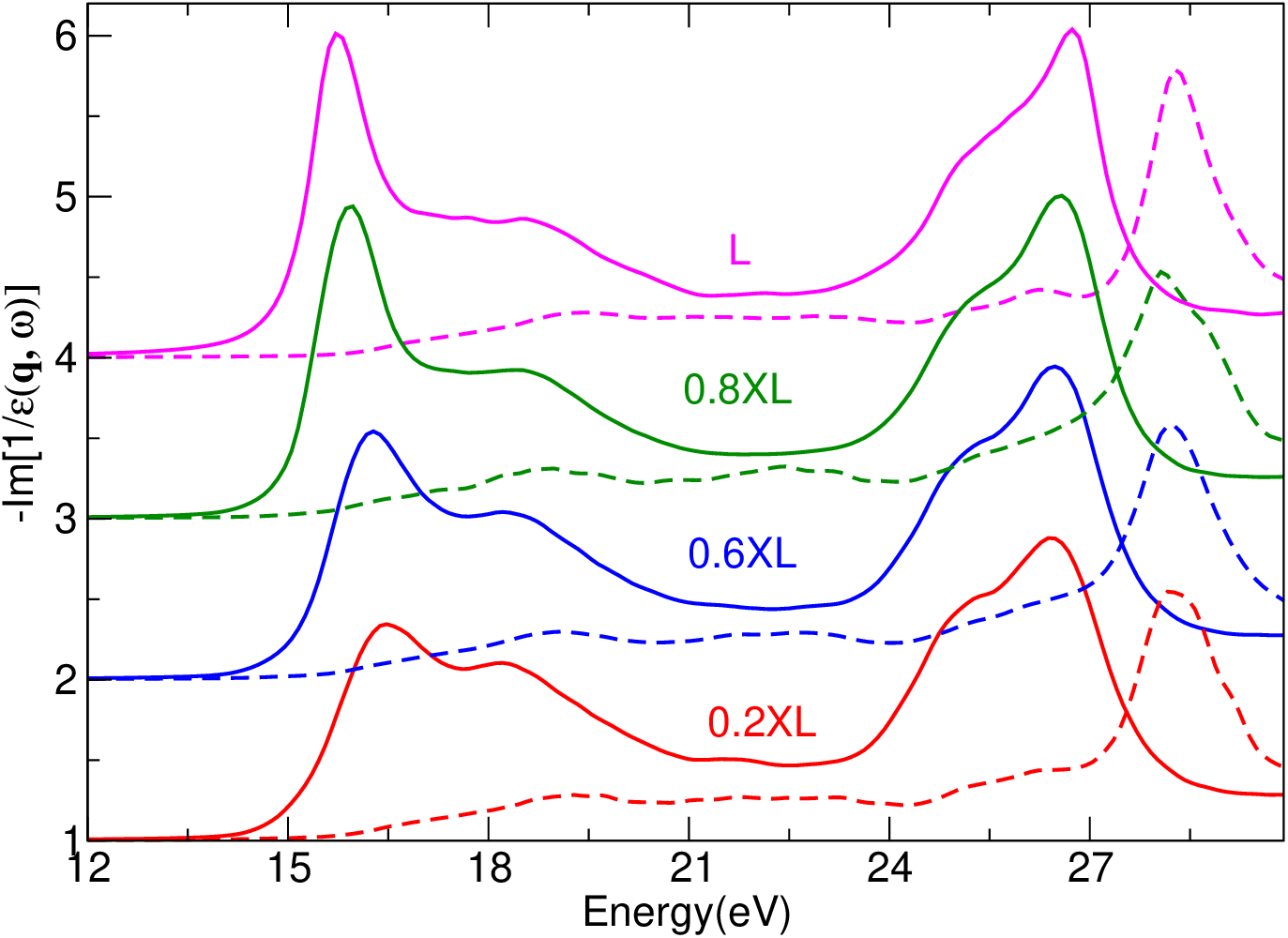}
\end{tabular}
\caption{Electron energy loss spectrum ($-Im[\varepsilon_{00}^{-1}({\bf q},\omega)]$)in arbitrary units
as a function of energy in eV for LiF. Results obtained using the bootstrap kernel are shown with full lines and
the results obtained using the RPA are shown with dashed lines. 
Upper panel shows results in the $\Gamma-L$ direction and lower panel in the $X-L$ direction. 
Results for various values of ${\bf q}$ are shifted vertically for clarity. 
\label{eels}}
\end{figure}
$\varepsilon_{00}^{-1}({\bf q},\omega)$ is an important quantity, not just because it can be directly compared to 
experiments, but also because it is required for an accurate determination of the screened Coulomb interaction, $W=\varepsilon^{-1}v$. 
In Hedin's theoretical foundation of many-body perturbation theory the effective interaction, $W$, is a crucial concept. 
It is needed as an essential ingredient
if one wishes to sum Feynman diagrams to obtain the Green's function of a system\cite{hanke78,onida02}.
In Fig. \ref{eels} we present results for the EELS of LiF as a function of ${\bf q}$ and frequency.
The EELS changes dramatically as a function of ${\bf q}$; the peak at 16 eV gains height on moving in the $\Gamma-L$ direction 
(upper panel), while at the same time the peak at 25 eV diminishes in magnitude and broadens. This leads to
the  EELS for ${\bf q} \rightarrow 0$ being significantly different from that at ${\bf q}=[0.5, 0.5, 0.5]$. 
A similar sharpening  of the peak at 16 eV is also seen in the $L-X$ direction (lower panel);
A broad double peak at $\sim$18 eV changes to one sharp peak at 16 eV and a small shoulder at $\sim$20 eV. In the $L-X$ 
direction (unlike the $\Gamma-L$) the peak at 25 eV stays almost unchanged as a function of ${\bf q}$. 
Similar strong changes in EELS as a function of ${\bf q}$, in the $\Gamma-U$ direction for LiF, have been reported before 
in Ref. [\onlinecite{marini04}].

The RPA results (also shown in Fig. \ref{eels}) not only miss the low energy excitonic effects\cite{marini04}, 
but also show very little variation as a function of ${\bf q}$ in both the $\Gamma-L$ and $X-L$ directions. Most many-body 
calculations use RPA dielectric functions for screening the Coulomb potential. However, it is clear from the present study 
that RPA dielectric functions can wildly differfrom those which properly include electron-hole physics. TDDFT with the bootstrap kernel  
is a computationally efficient method to accurately determine $\varepsilon^{-1}({\bf q},\omega)$ for use in
many-body perturbation theory\cite{marini04}.

To summarize, in the present work it is demonstrated that results obtained 
using the bootstrap kernel for finite values of
${\bf q}$ have the same accuracy as those in  the ${\bf q} \rightarrow 0$ limit. 
We use this kernel to make the predictions that in the prototypical materials LiF and Ar
excitonic effects are enhanced {\em away} from $\Gamma$. The reason for this enhancement is attributed to
the interaction between the exciton and other high energy inter-band transitions. 
This observation is reinforced by the application of an external electric field and noting an inverse proportionality
in strengths between the inter-band and excitonic peaks. 
It is further demonstrated that the EELS also changes dramatically as a function of ${\bf q}$. 
This strong change in EELS is missing within the RPA, and hence it is highly desirable to use a TDDFT dielectric
function to screen the Coulomb interaction as input for many-body perturbation theory.  

AR acknowledges funding from European Research Council Advanced Grant DYNamo (ERC-2010-AdG -Proposal No. 267374), 
Spain (FIS2010-21282-C02-01 and PIB2010US-00652), Grupos Consolidados UPV/EHU del Gobierno Vasco (IT-319-07) and 
ACI-Promociona (ACI2009-1036).

\end{document}